\documentclass{aa}  

\usepackage{amsmath}
\usepackage{natbib}
\usepackage{units}
\usepackage{color}

\usepackage{graphicx}
\usepackage{txfonts}

\newcommand{\beq}{\begin{equation}}
\newcommand{\eeq}{\end{equation}}

\begin{document}
	\title{Evolution of circumbinary planets around eccentric binaries: The case of Kepler-34}
	\author{Wilhelm Kley
 		\inst{1}
	\and
		Nader Haghighipour
 		\inst{2,3}
	}

	\institute{
	Institut f\"ur Astronomie und Astrophysik, Universit\"at T\"ubingen, Auf der Morgenstelle 10, D-72076 T\"ubingen, Germany. \\
	\email{wilhelm.kley@uni-tuebingen.de}
	\and
        Institute for Astronomy, University of Hawaii-Manoa, Honolulu, HI 96825, USA. \\
	\email{nader@ifa.hawaii.edu}
	\and
        Institut f\"ur Theoretische Astrophysik, Heidelberg, Germany \\
	}

\authorrunning{Kley \& Haghighipour}

 	\date{Received ; accepted }

\abstract
        {The existence of planets orbiting a central binary star system immediately 
        raises questions regarding their formation and dynamical evolution. Recent discoveries of circumbinary planets by
        the {\it Kepler} space telescope has shown that some of these planets reside close to the dynamical stability limit
        where the strong perturbations induced by the binary makes it very difficult to form planets in situ.
      For binary systems with nearly circular orbits, such as Kepler-38, the observed proximity of planetary orbits to the
      stability limit can be understood by an evolutionary process in which planets form farther out in the calmer environment 
      of the disk and migrate inward to their observed position.
%
      The Kepler-34 system is different from other systems as it has a high
      orbital eccentricity of 0.52. 
      Here, we analyse evolutionary scenarios for the planet observed around this system 
      using two-dimensional hydrodynamical simulations.
%
      The highly eccentric binary opens a wide inner hole in the disk which is also eccentric, and displays
      a slow prograde precession.
      As a result of the large, eccentric inner gap, an embedded planet settles in a final equilibrium position that lies 
      beyond the observed location of Kepler-34 b, but has the correct eccentricity. In this configuration the planetary orbit 
      is aligned with the disk in a state of apsidal corotation.
%
      To account for the closer orbit of Kepler-34 b to the central binary, we considered a two-planet scenario and examined the
      evolution of the system through joint inward migration and capture into mean-motion resonances. When the inner
      planet orbits inside the gap of the disk, planet-planet scattering ensues. While often
      one object is thrown into a large, highly eccentric orbit, at times the system is left with a planet close
      to the observed orbit, suggesting that Kepler 34 might have had two circumbinary planets where one might have
      been scattered out of the system or into an orbit where it did not transit the central binary during the operation of
      {\it Kepler}.
      }
   \maketitle
%

\section{Introduction}


With the success of the {\it Kepler} space telescope in the past few years in detecting several planets around main sequence
binaries, the formation and dynamical evolution of these objects are now among the most outstanding
questions in exoplanetary science. A survey of the currently known {\it Kepler}’s circumbinary planetary
systems\footnote{Currently known main sequence binaries with circumbinary planets are: 
\object{Kepler-16} \citep{2011Sci...333.1602D}, \object{Kepler-34} and 35 \citep{2012Natur.481..475W}, 
\object{Kepler-38} \citep{2012ApJ...758...87O}, \object{Kepler-47} \citep{2012Sci...337.1511O},
\object{Kepler-64} \citep{2013ApJ...768..127S} and \object{Kepler-413} \citep{2014ApJ...784...14K}.} 
indicates that, as expected, in all these systems, 
the binaries are close with orbital periods ranging from 7 to 41 days. The planets in these systems are
Saturn-sized or slightly smaller and revolve around their host binaries in orbits with periods from
50 to 300 days. Also, dynamical analyses of these objects have indicated that several of these planets 
are close to the boundary of orbital stability
\citep[see][]{1986A&A...167..379D,1999AJ....117..621H},
and they are all between two major $n:1$ mean-motion resonances (MMR).

The proximity of the orbits of some of the circumbinary planets (CBPs) to the stability limit raised the question that 
whether planet formation can proceed efficiently in these regions. Several attempts were made to answer this question
\citep{2012ApJ...761L...7M, 2012ApJ...754L..16P, 2013A&A...553A..71M, 2013MNRAS.435.2328D,2013ApJ...764L..16R,2014ApJ...782L..11L,2015arXiv150303876B}.
However, in situ formation appears not to be possible. The lack of success in growing planetesimals to larger sizes in the
vicinity of the stability boundary, combined with the fact that, as indicated by observations, the orbits of the
currently known circumbinary planets are almost all coplanar with the orbit of their host binaries, strongly suggested
that these planets formed at large distances (where the perturbing effect of the binary
on the disk is negligible and planet formation can proceed in the same way as around single stars) and
ended up in their current orbits either through planet migration (as a result of planet-disk interaction),
planet-planet scattering, or a combination of both \citep{2007A&A...472..993P, 2008A&A...483..633P}. 

The first study of planet migration in circumbinary disks was carried out by \citet{2003MNRAS.345..233N} who
showed that in low eccentricity binaries ($e_\text{bin} \leq 0.2$), massive, jovian-type bodies will be captured 
in a 4:1 MMR with the binary and when the binary is more eccentric, these planets
may be captured in stable orbits farther out \citep[see also][]{2008A&A...483..633P}. In a subsequent study,
\citet{2007A&A...472..993P} extended previous analyses to less massive planets (as low as 20 $M_\text{Earth}$)
and showed that migrating circumbinary planets may stop near the edge of the inner disk cavity. They suggested that 
circumbinary planets should be predominantly found in that area. Because in low eccentric binaries,
the inner edge of the disk cavity, which is due to the tidal effect of the binary on the disk material, almost coincides with the 
boundary of orbital stability, the predictions made by these authors agreed with the orbital architecture of the first few 
{\it Kepler} circumbinary planets. 
More general cases in which accretion and multiple planets were also considered were later studied by the same authors
\citep{2008A&A...478..939P,2008A&A...483..633P}.

The earliest attempt to explain the orbital architecture of Kepler CBPs using planet migration was made by \citet{2013A&A...556A.134P}.
Considering a locally isothermal disk, where the disk thermodynamics is modeled by prescribing a given temperature profile, 
and assuming a closed boundary condition at the edge of the cavity (i.e., no mass accretion onto the central binary), these
authors applied their model to the systems of Kepler 16, 34, and 35 and showed that the planets in these systems can migrate 
and reside in a stable orbit. However, the values of the semimajor axes of the final orbits of planets in their models did 
not agree well with their observed values. In a recent paper (Kley \& Haghighipour 2014), we showed that the limitation
of the models by \citet{2013A&A...556A.134P} are due to their two simplifying assumptions: considering disk to be locally 
isothermal and have a close boundary condition. A more realistic model requires the radiative effects to be taken into
account and the disk material to flow through the inner edge of the disk into the disk cavity. The latter has also been
shown in numerical simulations by \citet{1996ApJ...467L..77A} and \citet{2002A&A...387..550G} where the results indicate that
despite the appearance of a cavity in the center of a circumbinary disk, material can still flow inside and onto the central 
binary. 

In \citet{2014A&A...564A..72K}, we presented an improved and extended disk model which included a detailed balance of viscous 
heating and radiative cooling from the surface of the disk \citep{2003ApJ...599..548D}, as well as additional radiative diffusion 
in the plane of the disk \citep{2008A&A...487L...9K}. For a planet embedded in the disk, this improved thermodynamics can have 
dramatic effects on the planet orbital dynamics such that for low-mass planets, it can even reverse the direction of migration 
\citep{2008A&A...487L...9K,2009A&A...506..971K}. In addition, we considered an open boundary condition and allowed free in-flow 
of material from the disk into the central cavity. This enabled us to construct models with net mass in-flow through the disk. 
We showed that when our model is applied to circumbinary planetary systems with low eccentricity
or circular binaries (e.g., Kepler 38), it removes the discrepancy observed in the model by \citet{2013A&A...556A.134P},
and allows the planet to reside in close proximity to the boundary of orbital stability, between two $n:1$ MMRs. 

In our previous study, we considered the system of Kepler-38 as this system represents one of the CBP systems with an almost
circular binary. As mentioned above, in such systems, the boundary of the stability of planetary orbits coincides with the 
outer edge of the disk inner cavity. The latter raises the question that how the results will change if the binary is
eccentric. In a circumbinary disk around an eccentric binary, the inner cavity will be more eccentric. Because the inner
cavity is developed due to the tidal effect of the binary, its outer edge will no longer coincide with the stability boundary
as the latter also will change with the binary eccentricity. In this paper, we address these issues by applying our disk model
to the circumbinary system of Kepler 34. The central binary in this system has an eccentricity of 0.52 and the orbital
eccentricity of its circumbinary planet is 0.18. It has been suggested that the formation history of this system
may have been different from the more circular cases such as Kepler-16 and Kepler-38 \citep{2015ApJ...802...94G}. 

This paper is organized as follows: In the next section we briefly describe our hydrodynamic model.
In section 3, we present the structure of the disk without an embedded planet. In section 4, the results on 
the migration of a single planet are presented which is followed in section 5 by a study on the dynamics of 
a pair of embedded planets. In section 6, we summarize and discuss our results.

\section{The hydrodynamic model}
As mentioned in the introduction, the target of our investigation is the system of Kepler-34,
and we taylor our simulations to that particular system. Our overall methodology is identical to that used in our 
previous study of the Kepler-38 system \citep{2014A&A...564A..72K}, and we give here only a short outline. 

To model the evolution of a disk around a binary star, we
assume that the disk is vertically thin and the system is co-planar, an assumption well justified by the
observations of {\it Kepler} circumbinary planets.
We then perform two-dimensional (2D) hydrodynamical simulations in the plane of the binary.
To simulate the evolution of the disk, the viscous hydrodynamic equations are solved in a polar coordinate system 
($r,\phi$) with its origin at the center of mass of the binary. 

When using the polar coordinate system, there remains a hole in the center in which the binary orbits. 
As pointed out in previous studies \citep{2013A&A...556A.134P,2014A&A...564A..72K}, the location of the inner boundary
of the grid can affect the outcome of the simulations.
Given that the semi-major axis of the central binary of Kepler-34 is $a_\text{bin}=0.23$AU, we consider 
the radial extent of the disk ($r$) to be from 0.34 AU to 5.0 AU. Accordingly, as suggested by studies mentioned above,
the inner radius of the grid will be about 1.5 $a_\text{bin}$. The polar angle $\phi$ varies in an entire annulus of 
$[0,360^\circ]$. This domain is covered by a grid of $256 \times 256$ grid cells that are centrally condensed in the 
radial direction and equidistant in azimuth. In all models, we evolve the vertically integrated hydrodynamical equations 
for the surface density $\Sigma$ and the velocity components ($u_r, u_\phi$). 

When simulating locally isothermal disks, we do not evolve the energy equation and instead use an isothermal 
equation of state for the pressure. When considering radiative models,
a vertically averaged energy equation is used which evolves the temperature of the midplane 
\citep{2012A&A...539A..18M}. Radiative effects are included in two ways. 
First, a cooling term is considered to account for the radiative loss from the disk surface
\citep{2003ApJ...599..548D}. Second, we include diffusive radiative transport in the midplane of the disk 
using flux-limited diffusion \citep{2008A&A...487L...9K,2013A&A...560A..40M}.
In the radiative simulations, the full, vertically averaged dissipation is taken into account \citep{2012A&A...539A..18M}.

To calculate the necessary height of the disk $H$ at a position $\mathbf r$, following
\citet{2002A&A...387..550G} and \citet{2014A&A...564A..72K}, the individual distances from the two
central stars are taken into account.
The equation of state of the gas in the disk is given by the ideal gas law using a mean molecular weight of $\mu =2.35$ (in atomic mass units)
and an adiabatic exponent of $\gamma = 1.4$.
For the shear viscosity we use the $\alpha$-parametrization with $\alpha = 0.01$ or $0.004$ depending on the model,
and we set the bulk viscosity to zero.
For the Rosseland opacity, we use analytic formula as provided by \citet{1985prpl.conf..981L}
and the flux-limiter as in \citet{1989A&A...208...98K}. 

The calculation of the gravitational force acting by the binary and planets on the disk is done in the same
way as in \citet{2014A&A...564A..72K}. Here we use for the planet, a smoothing length of $0.6 H$ where the local scale height
takes both stars of the binary into account. In some of the simulations, we add a second planet. 
The integration of the hydrodynamical equations (explicit/implicit 2nd order scheme) and the N-body integrator 
(4th order Runge Kutta) are performed again identically to \citet{2014A&A...564A..72K}.
To calculate the force from the disk acting on the planet, we exclude parts of the Hill sphere of the planet
using a tapering function as given in \citet{2014A&A...564A..72K}.
We calculate the orbital parameters of the planet  
using Jacobian coordinates, assuming the planet orbits a star with mass $M_\text{bin} = M_1 + M_2$, at the binary
barycenter.

\subsection{Initial setup and boundary conditions}
\label{subsec:setup}
\begin{table}[t]
\caption{The binary parameter and the observed planetary parameter of the Kepler-34 system
 used in the simulations.
 The mass of the primary star is $1.05 M_\odot$. The values were 
 taken from \citet{2012Natur.481..475W}.
 \label{tab:kepler34}
}
a) Binary Parameter \\
\medskip
\begin{tabular}{l|l|l|l} 
\hline
 Mass ratio   &  Period  &  $a_\text{bin}$   &   $e_\text{bin}$   \\  
   $q = M_2/M_1$  &  [days]   &    [AU]       &           \\
\hline
  0.97          &   28  &  0.23    & 0.52    \\
\hline
\end{tabular} 

\medskip

b) Planet Parameter \\
\medskip
\begin{tabular}{l|l|l|l}
\hline
 Mass   &  Period  &  $a_\text{p}$   &   $e_\text{p}$   \\
   $M_\text{Jup}$  &  [days]   &    [AU]       &           \\
\hline
  0.22          &   288  &  1.09    &  0.18     \\
\hline
\end{tabular}

\end{table}

In our models, the disk initial surface density is chosen to have a $\Sigma (r) = \Sigma_0 \, r^{-1/2}$ profile where $r$ is the
distance from the center of mass of the binary. For this reference surface density, following our
earlier results \citep{2014A&A...564A..72K}, we chose two different values, a higher mass model with $\Sigma_0 = 3000$g/cm$^2$ 
and a lower mass model with a quarter of this value, $\Sigma_0/4 = 775$g/cm$^2$.
We chose the initial temperature of the disk to vary with $r$ as $T(r) \propto r^{-1}$ such that, 
assuming a central star of $M_\text{bin}$, the vertical thickness of the disk will always maintain the condition $H/r = 0.05$.
For the radiative model, the temperature profile is determined self-consistently.
The initial angular velocity of the disk at a distance $r$ from the barycenter is chosen to be equal to the Keplerian velocity at that
distance, and the radial velocity is set to zero.

The parameters of the central binary have been adopted from \citet{2012Natur.481..475W} and are quoted in in Table~\ref{tab:kepler34},
together with the planetary data.
Owing to the interaction of the binary with the disk (and planet), these parameters will slowly vary during a simulation.
However, during the course of our simulations the change is very small. 
Even for the longest runs that stretch over 10,000 yrs (equivalent to over $1.3 \cdot 10^5$ binary orbits) the relative change
of the semi-major axis and the eccentricity of the binary is $3 \cdot 10^{-3}$ and  $6 \cdot 10^{-3}$, respectively.
For the models with an embedded planet, we consider the planet to have a mass of $m_\text{p} = 2.1 \times 10^{-4} M_1$
which is equivalent to $m_\text{p} = 0.22 M_\text{Jup}$ or about 73 Earth masses (see Table~\ref{tab:kepler34}).
At the beginning of each simulation, we start the binary at its periastron. 
 
The boundary conditions of the simulations are constructed such that at the outer boundary of the disk, $r_{max}=5.0$ AU,
the surface density remains constant and equal to its initial value.
This is achieved by using a damping boundary condition where the density is relaxed toward
its initial value, $\Sigma(r_{max})$, and the radial velocity is damped toward zero, using the
procedure specified in \citet{2006MNRAS.370..529D}. 
The angular velocity at the outer boundary is also kept equal to its initial Keplerian value.
For the temperature, we use a reflecting condition such that there will be no artificial radiative flux through the 
outer boundary for the radiative model. These conditions at the outer boundary lead to a disk with zero eccentricity 
at $r_{max}$. Hence, $r_{max}$ has to be large enough such that the inner regions are not influenced.

At the inner boundary $(r_{min})$, we consider a boundary condition such that the in-flow of disk material onto the binary is allowed.
This means, for the radial boundary grid cells at $r_{min}$, we choose a zero-gradient mass out-flow condition, 
where the material can freely leave the grid and flow onto the binary. No mass in-flow into a grid is allowed at $r_{min}$.
The zero-gradient condition is also applied to the angular velocity of the material since because of the effect of the binary,
no well-defined Keplerian velocity can be found that could be used otherwise. This zero-gradient condition for the
angular velocity implies a physically more realistic {\it zero-torque} boundary.

With these boundary conditions, the disk can reach a quasi-stationary state in which there will be a
constant mass flow through the disk  \citep{2014A&A...564A..72K}.

\begin{table}
\caption{Parameters of the locally isothermal disk reference model without a planet.
 \label{tab:modref}
}
\begin{tabular}{l|l}
 Parameter   &  Value  \\             
\hline
 Surface density at 1 AU   &  3000 g/cc  \\
 Viscosity ($\alpha$-value)   &    0.01  \\
 $R_{min},R_{max}$  &    0.34 AU, 5.0 AU  \\
  fixed H/R       &   0.05  \\
\hline
\end{tabular}  
\end{table}

\begin{figure}
\center
\includegraphics[width=0.45\textwidth]{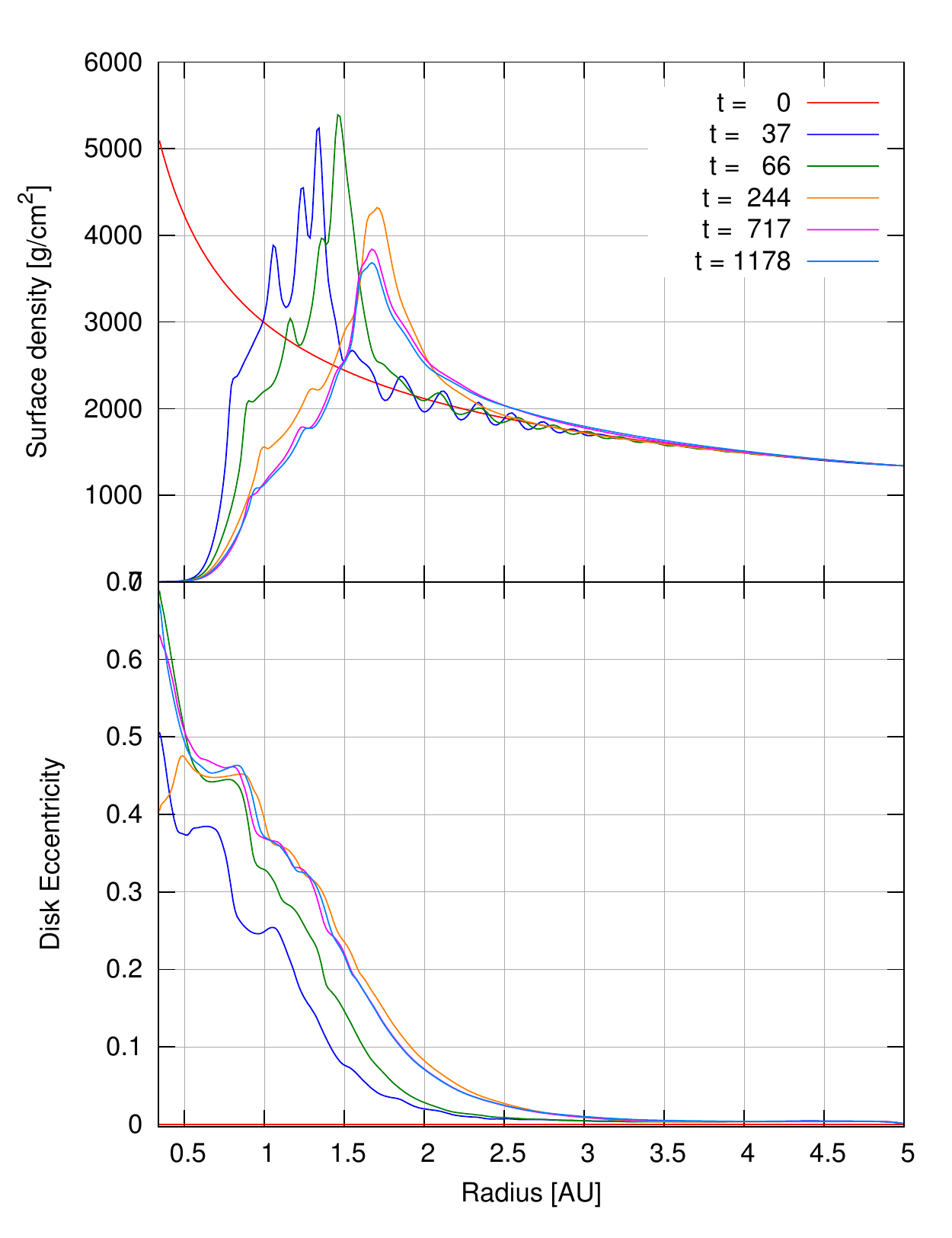}
\caption{Azimuthally averaged surface density (top) and disk eccentricity (bottom) for our locally isothermal reference model.
Shown here are the profiles at different evolutionary times (in years); the red curve denotes the initial
setup.
}
\label{fig:k34a-sigecc}
\end{figure}

\section{The structure of the circumbinary disk}
\label{sec:disk-structure}
Before we study the evolution of planets within the circumbinary disk around Kepler-34, we
analyse the disk's dynamics due to the effect of the central binary.
For this purpose, we simulated the dynamics of the disk in our reference model (see Table~\ref{tab:modref}) with a zero-mass planet.
In the simulations the inner boundary of the disk is open to mass flow into the central cavity, 
and at the outer boundary the density is relaxed to its initial values.
Starting from the initial setup mass flows out into the inner cavity and the binary's torques change the disk's density profile.
However, caused by the relaxation outer boundary condition it eventually settles into a new final quasi-stationary state, where
its mass remains constant and there is a constant mass accretion rate onto the central binary. This behaviour is similar to our
previous simulations for the Kepler-38 system \citep{2014A&A...564A..72K}.
We construct our disk model for a locally isothermal equation of state where we do not evolve the energy
equation but instead leave the temperature of the disk at its initial value.
This procedure has the advantage of making the simulations much faster
as no heating and cooling of the disk have to be considered.


\begin{figure}
\center
\includegraphics[width=0.45\textwidth]{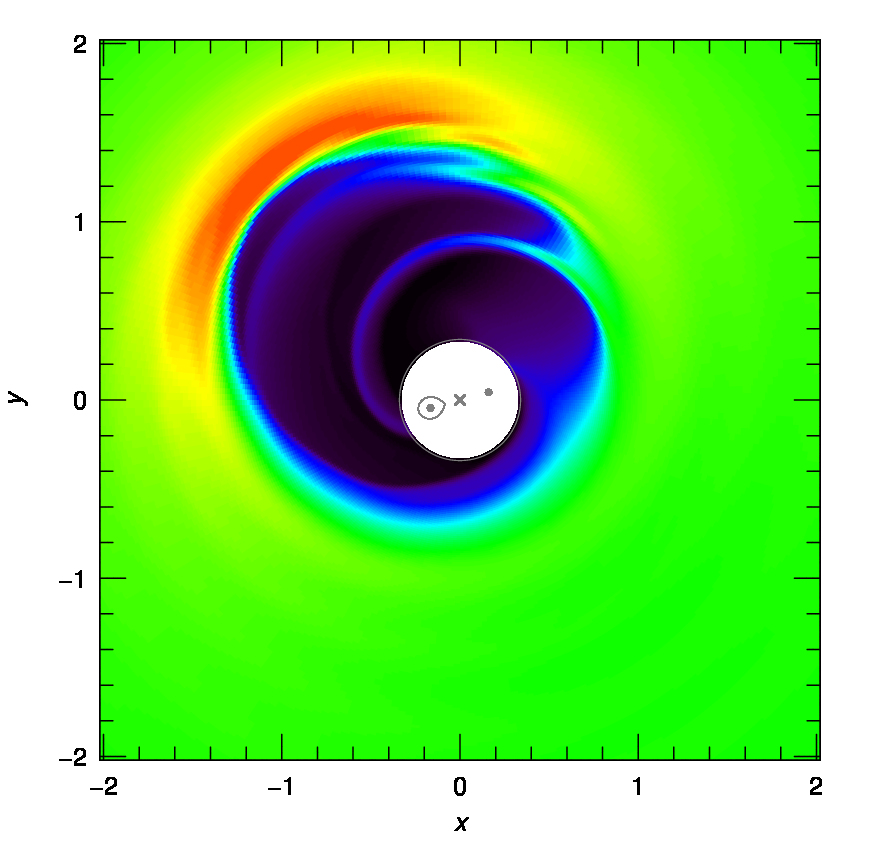} \\
\caption{Two-dimensional density structure of our isothermal reference disk model around the central binary star.
The graph shows only a local view around the binary. The entire computational grid, however, extends from $r=0.34$ AU to $r=5.0$ AU.
The white inner region lies inside the computational domain and is not covered by the grid.
The positions of the primary and secondary stars are indicated by the gray dots. The Roche lobe of
the secondary is also shown. The central cross marks the origin of the coordinate system which coincides with
the center of mass of the binary.
}
\label{fig:2D-k34a}
\end{figure}

\subsection{Disk structure}
\label{subsec:diskstruct}

During time, the surface density slowly evolves away from its initial profile
until it settles in a new equilibrium state. 
This is shown in the upper panel of Fig.~\ref{fig:k34a-sigecc} where the azimuthally
averaged radial surface density profile is shown at different evolutionary times. 
The equilibration time takes only a few hundred years. After this time,
the surface density profiles will no longer change considerably. Because of the tidal action of the binary 
on the disk, a central gap is formed with a surface density many orders of magnitude lower than inside the disk. 
The inner edge of the disk, which we can define, very approximately, as that radius at which
the surface density is about half the maximum value, lies here at around $r \approx 1.25$AU.
This is about 5 times larger than $a_\text{bin}$ and as such larger than
the stability region of massless particle trajectories around binary stars as given by \citet{1994ApJ...421..651A}.
Our result is in good agreement with \citet{2013A&A...556A.134P} and our own results on Kepler-38 
\citep{2014A&A...564A..72K}. The larger value of the inner gap is caused by the
high eccentricity of the binary disk in this case. 
Outside of about $r=3$AU, the surface density profile remains at the initial value.

As was shown by \citet{2013A&A...556A.134P}, circumbinary disks can attain significant eccentricities.
We calculate the eccentricity of the disk by treating each grid cell as an individual particle with
a mass and velocity equal to the mass and velocity of the cell \citep{2006A&A...447..369K}. 
To calculate a radial dependence for the disk eccentricity, $e_\text{disk}(r)$, we average over the angular direction.
In our simulations, the disk eccentricity becomes very high, with $e_\text{disk}$ about 0.3 to 0.5, 
in the gap region of the disk, as shown in the lower panel of Fig.~\ref{fig:k34a-sigecc}.
In the inner, nearly evacuated regions, the disk eccentricity can be even higher.
Near the maximum of the density (at $r=1.7$), the eccentricity is around 0.15 and it drops slowly further out. 
At radial distances $r > 3.0$ AU, the disk eccentricity becomes lower than about $0.01$.

\begin{figure}
\center
\includegraphics[width=0.45\textwidth]{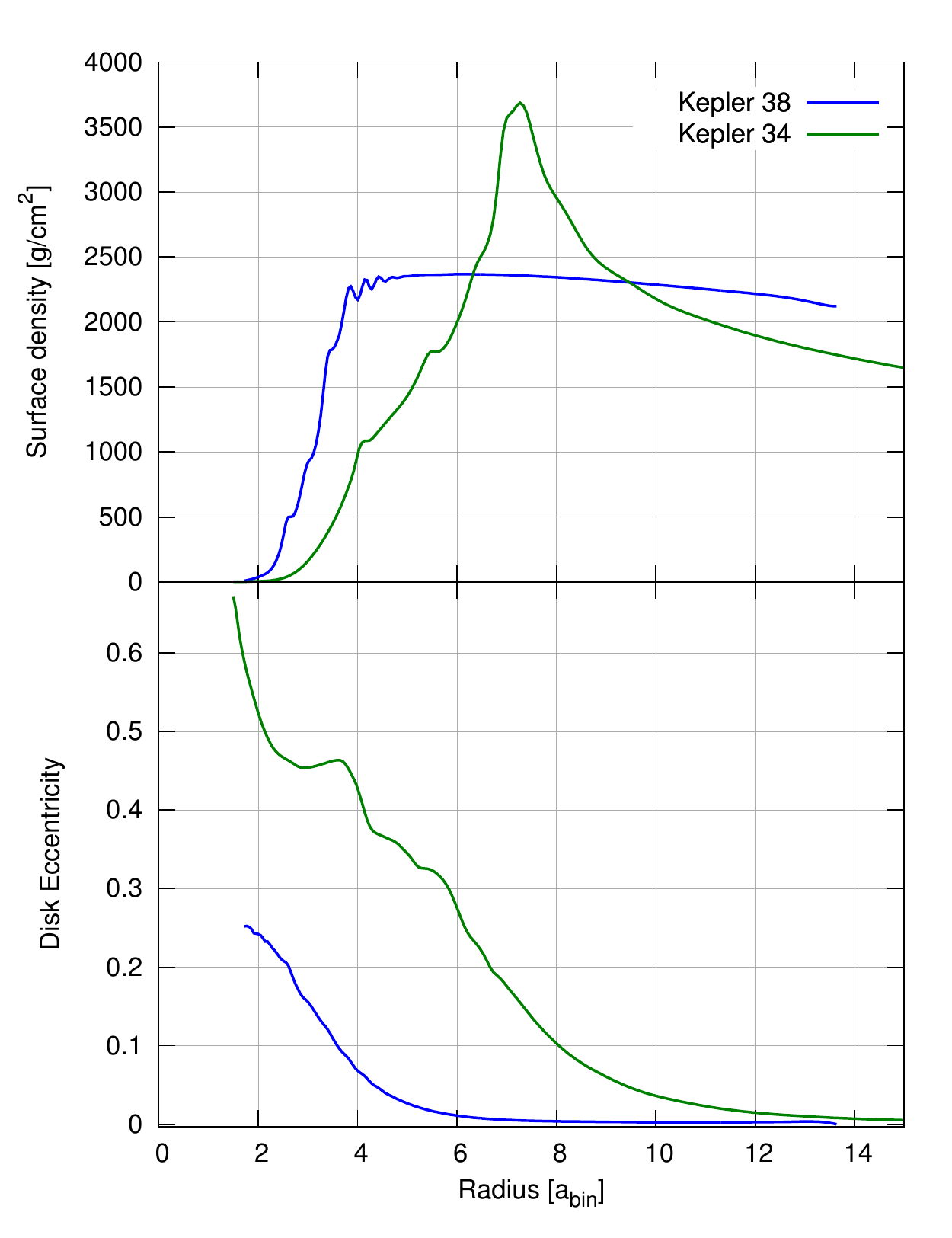}
\caption{Azimuthally averaged surface density (top) and eccentricity (bottom) for locally isothermal disks,
with $H/r=0.05$, around the two systems, Kepler-34 and Kepler-38.
The figure shows the profiles for the final equilibrium state without any embedded planet.
The radial coordinate has been scaled by the binary separation which is $0.23$AU for Kepler-34 and
$0.15$AU for Kepler-38.
The results for Kepler-38 have been taken from \citet{2014A&A...564A..72K}.
}
\label{fig:sigecc-34-38}
\end{figure}

In Fig.~\ref{fig:2D-k34a}, we show the 2D surface density distribution for the isothermal
disk models. We note that this figure shows only the inner part of the computational domain around the central binary. 
The Roche lobe of the secondary star is shown as well. As shown here, an eccentric central binary strongly perturbs
the disk and produces time varying patterns and a very large eccentric inner gap.
At the same time, the disk features a strong asymmetric maximum in the surface density.

To test the sensitivity of the results to changes in the inner boundary condition, we performed a comparison
model with a closed inner boundary and found no significant differences in the results concerning the gap structure.
The width of the gap and the disk eccentricity were the same, only the value of the maximum density was slightly higher 
in the closed boundary model.

In order to study the influence of the binary parameters on the disk structure, we compare in
Fig.~\ref{fig:sigecc-34-38} the averaged disk structure for the Kepler-34 system and the Kepler-38 system.
These two systems are different in mass ratio and eccentricity, where for the Kepler-38 system, $q=0.26, e=0.10$.
It is clear that the combination of a higher mass ratio together with a higher binary eccentricity creates a much
larger variation in the surface density profile and a higher disk eccentricity. This difference will have a profound effect
on the dynamical evolution of embedded planets as shown in section \ref{sec:withplanet}.

\begin{figure}
\center
\includegraphics[width=0.45\textwidth]{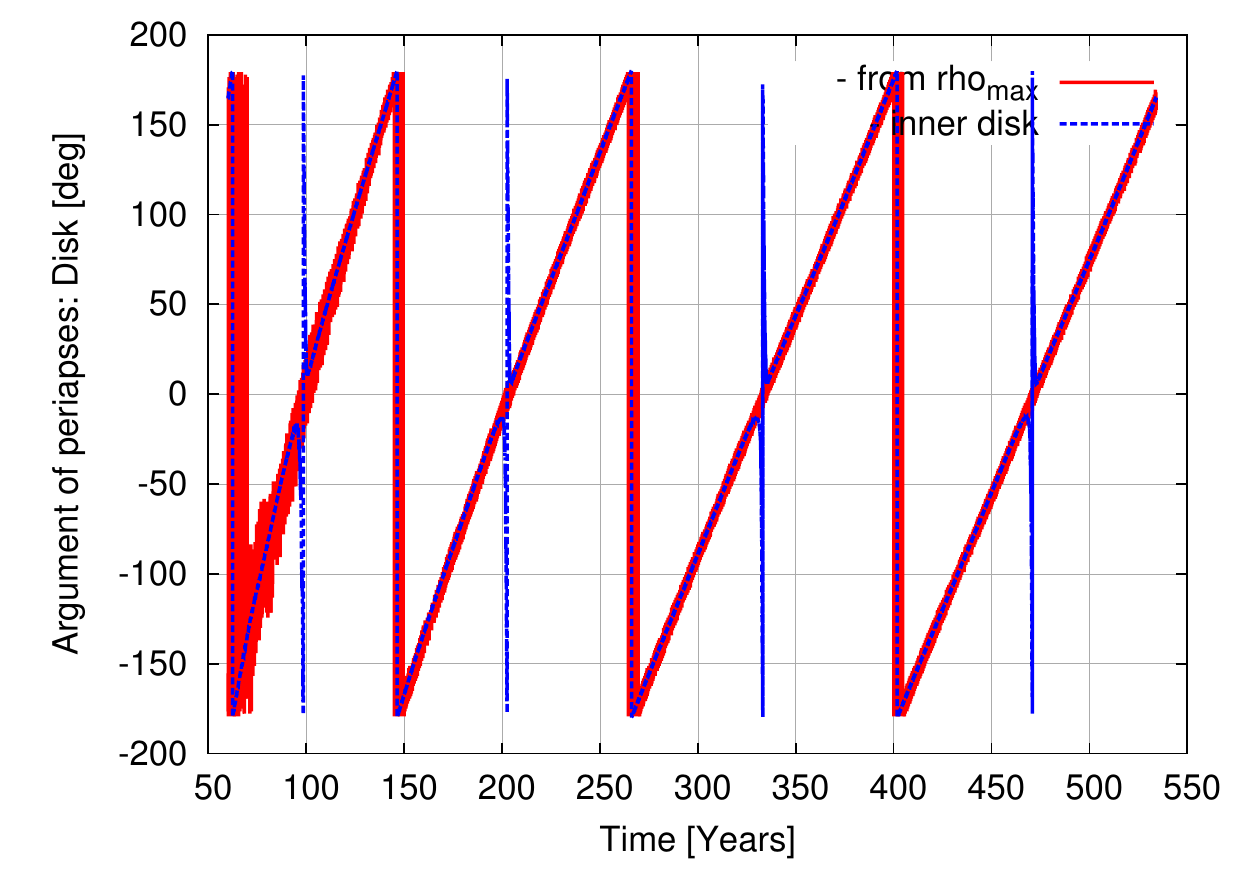} \\
\caption{The evolution of the argument of periapses of the disk with respect to the inertial frame.
Results of two different estimates are shown for comparison. 
Red corresponds to estimates using the maximum of the 2D density distribution (see Fig.~\ref{fig:2D-k34a}), and blue 
is for estimates using the mass weighted mean value of the inner disk.
}
\label{fig:k34h-omc1}
\end{figure}

\subsection{Disk precession}
\label{subsec:diskprec}
 
As outlined above, the surface distribution around the central binary becomes clearly eccentric.
To analyse the disk dynamics without a planet, we further analyzed the precession rate of this eccentric
mode. In principle, the argument of periapse of the disk, $\varpi_{\mathrm disk}$,
can be calculated in the same manner as the disk eccentricity,
as a mass-weighted average over the disk integrating over all annuli \citep{2006A&A...447..369K}. 
However, when calculating the integral over the entire disk from $r_{min}$ to $r_{max}$, the result
showed a nearly constant value for $\varpi_\mathrm{disk}$. However, animations
of the surface density distribution, as shown for a single snapshot in Fig.~\ref{fig:2D-k34a}, indicated a clear precession
of the inner disk regions. For that reason, we decided to restrict the range of integration
to the inner disk only, and used a radial range extending from $r_{min}=0.34$ to $r = 2$. As an additional indicator for the 
precession, we calculated the disk's line of periapses from the maximum of the density distribution in the disk.
The two methods resulted in similar results as shown in Fig.~\ref{fig:k34h-omc1} where the time evolution of
$\varpi_\mathrm{disk}$ is displayed for the reference model without an embedded planet.
\begin{figure}
\center
\includegraphics[width=0.45\textwidth]{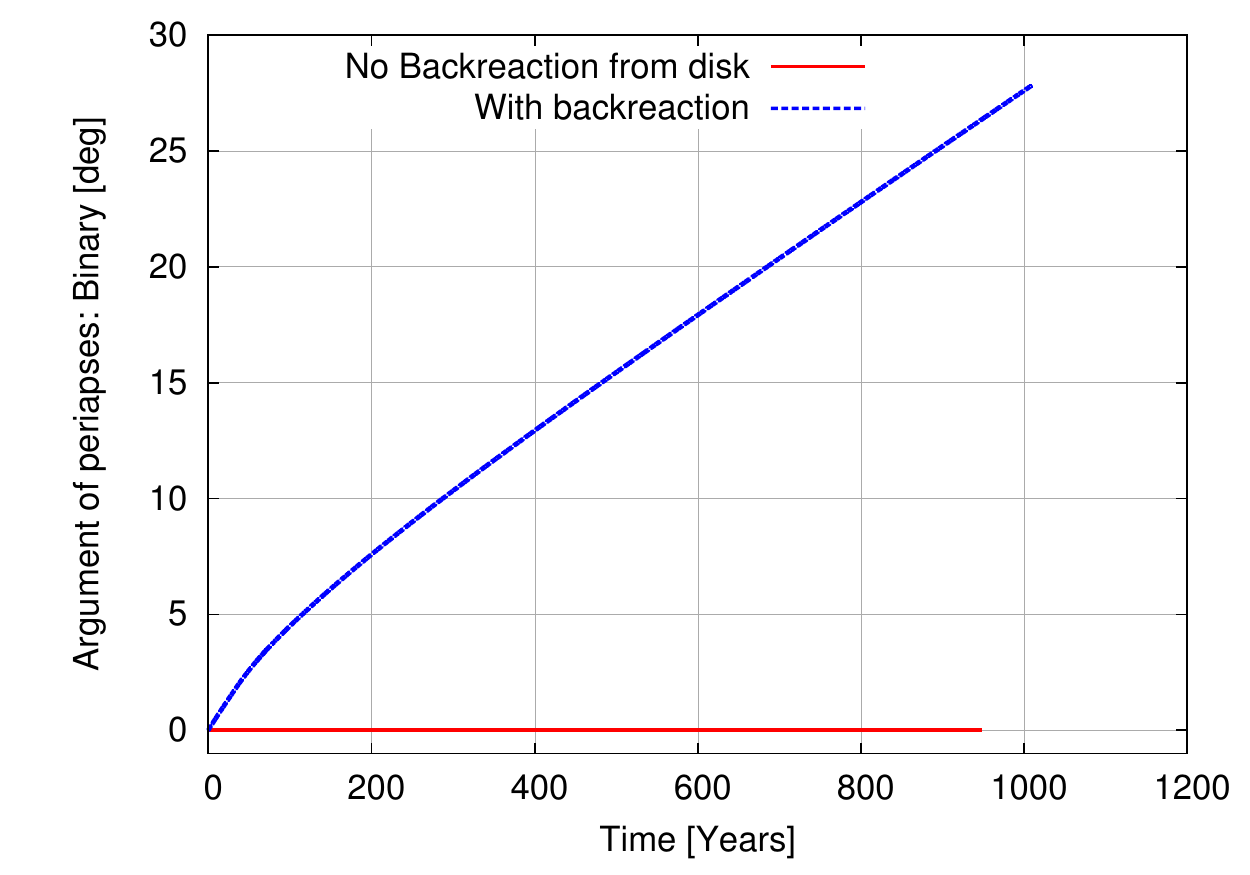} \\
\caption{The evolution of the argument of periapses  of the binary with respect to the inertial frame.
The red curve corresponds to the case where the backreaction of the disk onto the binary has been
articifially switched off such that the motion of the binary is not influenced by the presence of the disk.
The blue curve correponds to the reference model as shown in Fig.~\ref{fig:2D-k34a}.
}
\label{fig:k34h-omb}
\end{figure}
The above-mentioned averaging method gives strong fluctuations whenever the angle crosses zero because individual rings may
have values close to zero or close to 360$^\circ$. The density maximum method on the other hand is noisier due to
the disk dynamics which results in the 'thicker' curves. 
The overall agreement of the two methods confirms that the disk experiences a precession.
In the initial phase of the simulations the precession rate adjusts to the changing density profile
and in the final equilibrium state the disk experiences a precession with a rate of about $\dot{\varpi}_\mathrm{disk} \approx 3^\circ$/yr.
In comparision to the orbital period of the binary, $P_\mathrm{bin} = 28$ d, this indicates a very slow precession rate.
Similar slow precession rates have been found for disks around massive planets \citep{2006A&A...447..369K} as well as the
inner disks in close binary systems such as Cataclysmic binaries \citep{2008A&A...487..671K}. 
The reason that \citet{2013A&A...556A.134P} did not find precession of the disk in their models is probably related
to an integration over the global disk. 

In this paper, we do not investigate further details of our disk dynamics, but only briefly comment on the feedback 
the disk can have on the central binary. 
In Fig.~\ref{fig:k34h-omb}, we show the precession rate for the binary as induced by the disk. As expected, without the
disk feedback, the binary orbit does not precess. However, the inclusion of the disk's feedack leads to a very slow prograde
precession of about $\dot{\varpi}_\mathrm{bin} \approx 0.05^\circ$/yr in the orbit of the binary. Here, the total mass of the disk 
is about 0.015 $M_\odot$.  During the time that the orbit of the binary developed precession, the disk yielded a slow 
decrease of the orbit of the binary with a rate of about $4 \cdot 10^{-7}$AU/yr.
 
\begin{figure}
\center
\includegraphics[width=0.45\textwidth]{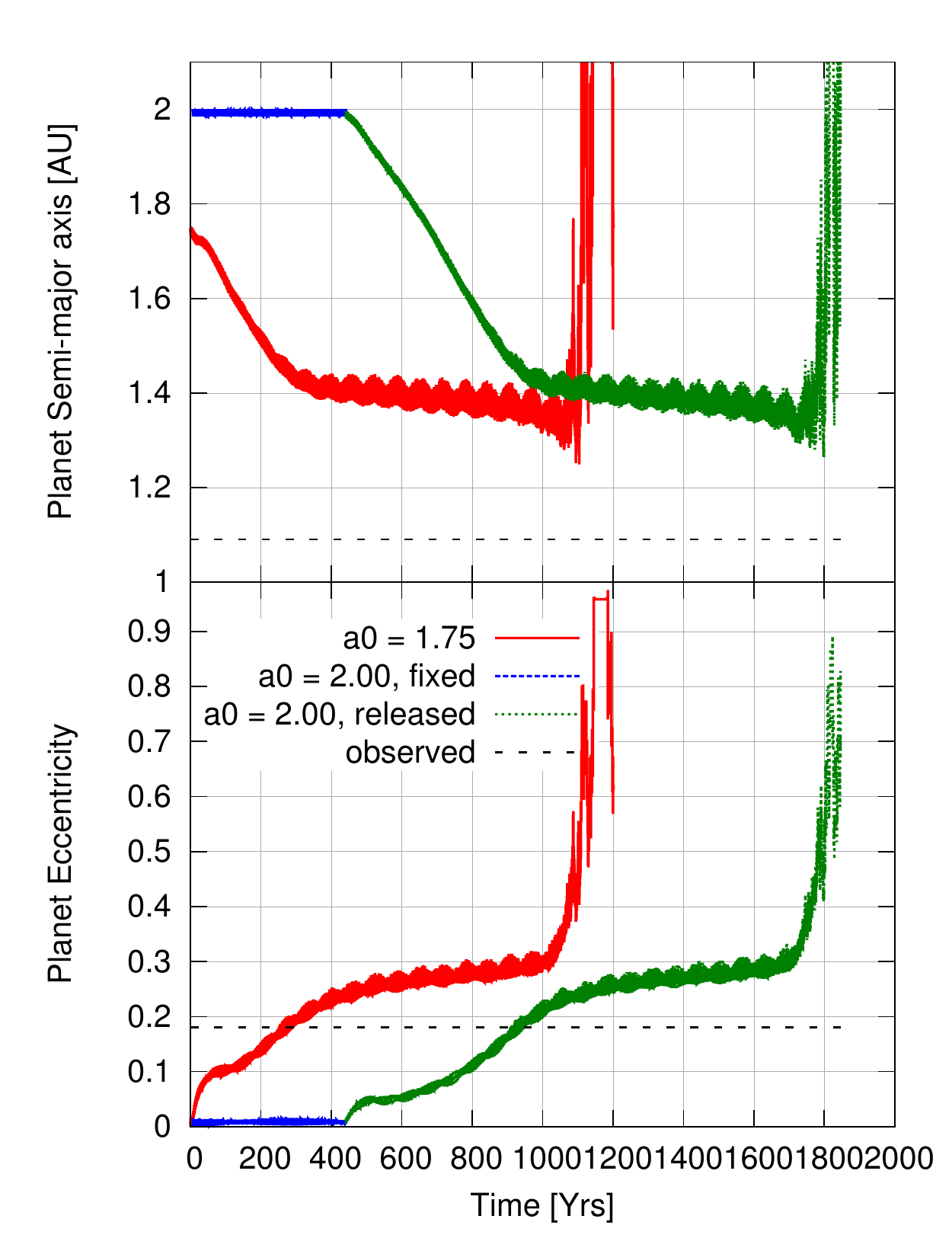} \\
\caption{The evolution of the semi-major axis (top) and eccentricity (bottom) of an embedded planet in an isothermal disk (reference model)
that was first brought into an equilibrium.
In the two simulations, planet was started at different distances from the center of mass of the binary. 
In the first simulation (shown in red), the planet started at a distance of $a_0 =  1.75$~AU, and in the second simulation (shown in green/blue)
it started as  $a_0 =  2.0$~AU.
In the simulation shown in red, the planet is immediately released and evolves with the disk. In the second simulation, the planet's orbit
is held constant during the first 600 yrs (in blue) and then released (green lines). 
In both simulations, planet migrates inward with a rate of about 0.1 AU /100 yrs. Both models result in unstable evolution
when the planet has reached a distance of about 1.35 AU. 
The dashed horizontal lines shows the observed semi-major axis and eccentricity of Kepler-34 b.
}
\label{fig:ap-ep-iso}
\end{figure}

\section{Planet migration}
\label{sec:withplanet}
In this section, we analyse the evolution of embedded planets in the disk around Kepler-34. 
Similar to our previous study on Kepler-38 \citep{2014A&A...564A..72K}, we first study migration for
our standard isothermal disk model because these are numerically faster. 
Subsequently, we study planets in radiative disks, as well.  

\subsection{Migration in the standard disk model}
\label{sec:isodiskplanet}
To study the evolution of the system with an embedded planet, we start our simulations with a planet initially placed at
different distances (semi-major axes, $a_0$) from the center of mass (barycenter) of the binary and in a circular orbit. 
The planet's evolution is determined by the gravitational action of the disk and the central binary.
As mentioned earlier, during the evolution of the planet, its orbital elements
are calculated using Jacobian coordinates with respect to the barycenter of the central binary.
The planet mass is fixed to 73 Earth masses (the observed value), and there is no accretion onto the planet.
As a reference, we present in Table~\ref{tab:kepler34}, the present orbital parameters of the planet Kepler-34 b,
as inferred from the observations.

Figure~\ref{fig:ap-ep-iso} shows the evolution of the planet through the disk for the reference model.
We present here the results of two simulations that were carried out for different initial conditions of the planet from the center of mass
of the binary. 
In the first model (red line), the planet is started directly after insertion. Initially, it rapidly migrates inward until 
it has reached a distance of about 1.4\,AU at around 300\,yrs into the evolution. After this, the inward migration proceeds at a lower pace
because the planet has reached the eccentric inner hole of the disk. Upon reaching a distance of 1.3\,AU 
the eccentricity of the planet has increased to about 0.3 and its orbit becomes unstable.
To examine whether this outcome might have been caused by a too small initial separation of the planet, so that the
system could not reach an equilibrium, we started a second model with the
planet embedded initially farther out at $r_0 = 2.0$AU. During the first 600 years, the semi-major axis of the planet was held constant
(blue lines in Fig.~\ref{fig:ap-ep-iso}) which gave enough time to the disk to adjust to the presence of the planet.
The planet was then released and it migrated due to the effect of the disk torque acting on it (green lines).
The planet drifted inward with the same speed as in the first model, about 0.1AU/100yrs, and the
orbit became unstable after the planet reached the same position as before.
This implies that the outcome of the migration process does not depend on the history
of the system and is determined solely by the physical parameters of the disk. 
In the following we analyze if reducing migration speed can stabilize the orbit. 
 
\begin{figure}
\center
\includegraphics[width=0.45\textwidth]{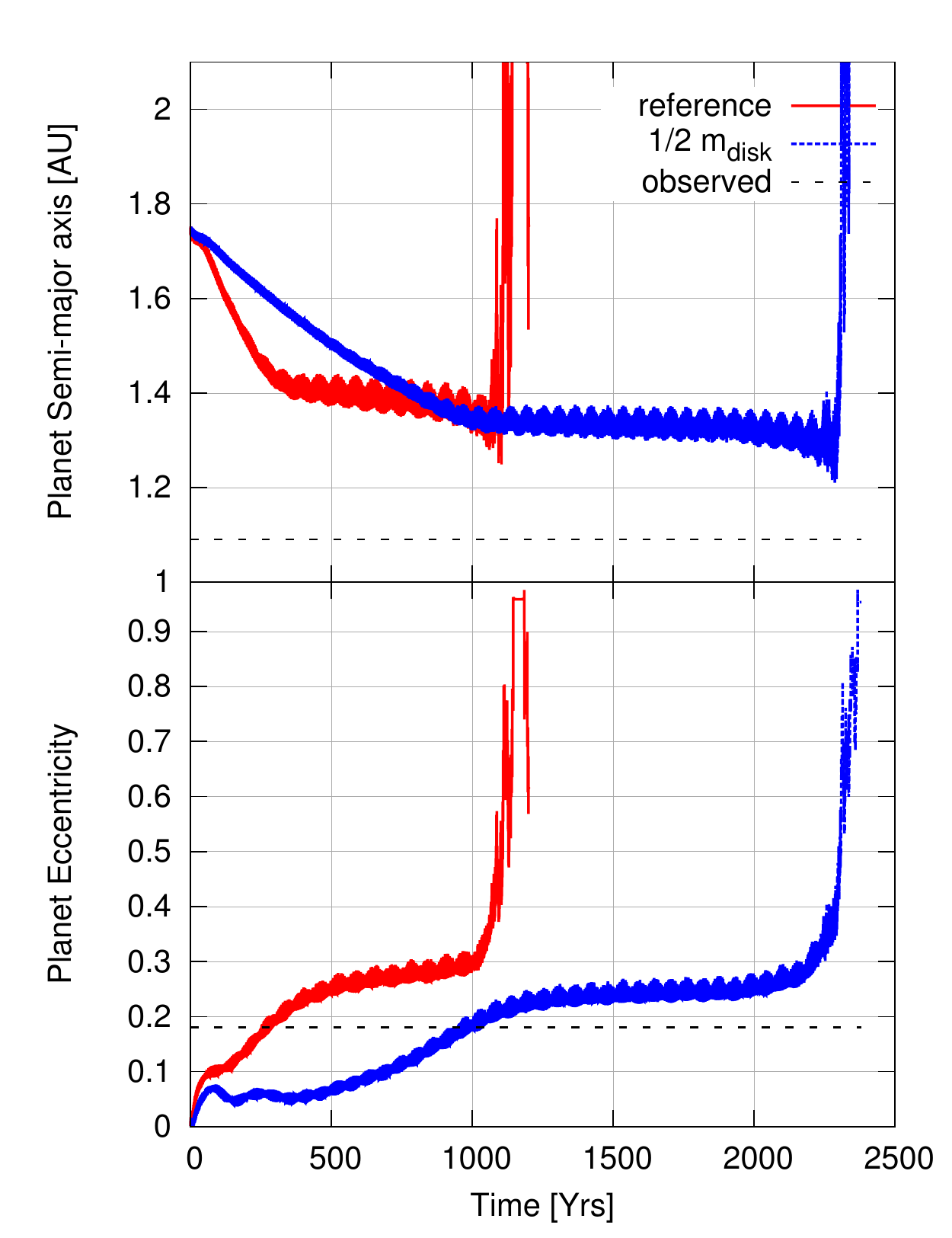} \\
\caption{The evolution of the semi-major axis (top) and eccentricity (bottom) of an embedded planet
in isothermal disk models with half the disk mass. The red curve refers to the original model and is identical to the one shown in
Fig.~\ref{fig:ap-ep-iso}.
}
\label{fig:k34c1-aep}
\end{figure}

\subsection{Planet migration in disks with lower mass}
\label{sec:redmassplanet}
To investigate whether it was the very rapid inward migration of the planet in our standard model that resulted
in its unstable evolution, we performed additional isothermal disk simulations with a reduced disk mass but otherwise identical
parameter. The results are shown in Fig.~\ref{fig:k34c1-aep}. Here, the red line refers to the standard model as shown with
the same color in Fig.~\ref{fig:ap-ep-iso} and the blue line shows a model with 1/2 the disk mass. The reduction of the disk mass resulted
in a slower inward migration with slightly reduced eccentricity, however, the model became unstable again.

\begin{figure}
\center
\includegraphics[width=0.45\textwidth]{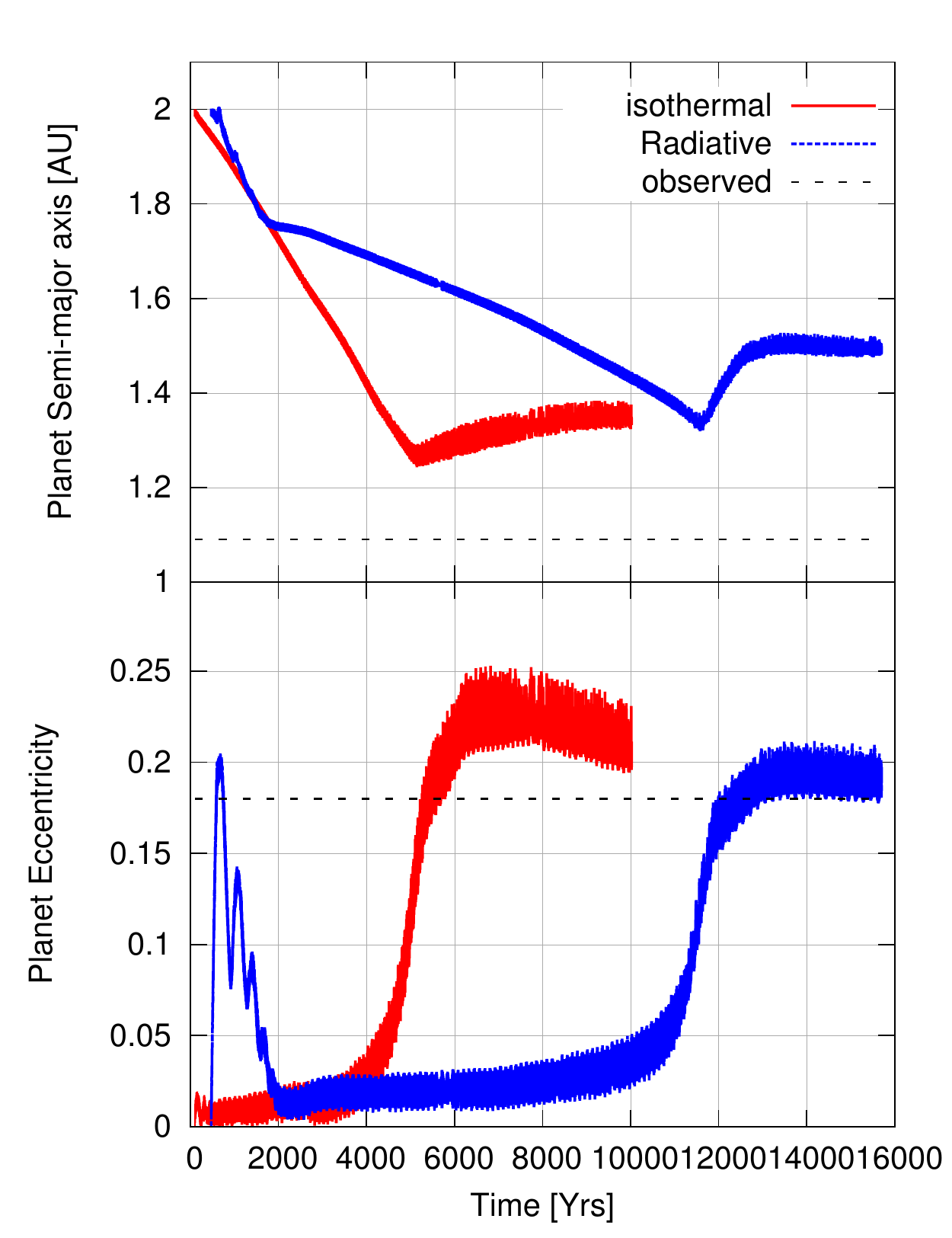} \\
\caption{The evolution of the semi-major axis (top) and eccentricity (bottom) of an embedded planet in 
disks with lower surface densities (1/4 of the reference model) and reduced viscosities ($\alpha=0.004$).
The red graph corresponds to the isothermal model and the blue refers to the radiative disk.
In the two simulations, the planet is started in a circular orbit at a distance of $a_0 =  2.0$~AU 
from the center of mass of the binary. 
In the simulation shown in red, the planet is included in the disk from the very beginning of the simulation
and it begins to migrate simultaneously with the disk evolution starting from its initial density profile.
In the simulation shown in blue, the disk is first relaxed to the radiative equilibrium and then the planet is embedded
in it.  The dashed horizontal lines refer to the observed parameter of the Kepler-34 planet.}
\label{fig:k34f1p-ap-ep}
\end{figure}

\subsection{Planet migration in disks with lower viscosity and with radiation}
\label{sec:redviscplanet}
In addition to the disk mass, we also varied the viscosity and thermodynamics of the disk to investigate
its influence on the migration process. Fig.~\ref{fig:k34f1p-ap-ep} shows the semi-major axis and eccentricity of the planet
embedded in a disk with a lower surface density (1/4 of the reference model) and a reduced disk viscosity ($\alpha=0.004$).
We compare a locally isothermal model (in red) with a radiative one (in blue). In the isothermal model, the planet migrates inward 
at a constant rate which is about five times slower than in the reference model. When the planet reaches a distance slightly 
inside of $r = 1.3$ AU, it reverses its direction and migrates outward until it finally settles at a distance of about 1.35 AU. 
In the radiative case, the planet migrates inward initially at a fast pace, but then it continues its inward migration much more 
slowly than the isothermal model. The overall evolution of the planet in this case is similar to the isothermal one. 
The final orbit of the planet is only slightly further away
at approximately 1.5\,AU. Given the similarity of the isothermal and radiative results, we decided to continue our study
in the rest of the paper by using primarily the isothermal approximation in order to cut
down on unnecessary computation time.

Th interesting outward migration of the planet near the end of its evolution can be understood
in terms of the interaction of the eccentric disk with the migrating planet. During the inward migration of the planet, 
the eccentricity of the planet’s orbit remains small. We noticed that when the planet
approaches the inner, more eccentric regions of the disk, the disk becomes more circular which allows further inward migration
of the planet.
Once the planet reaches the inner hole of the disk, the eccentricities of the disk and the planet increase again causing the
planet move into the disk, periodically. Subsequently, the planet turns around
and moves slightly outwards, as shown in Fig.~\ref{fig:k34f1p-ap-ep}. 

\begin{figure}
\center
\includegraphics[width=0.45\textwidth]{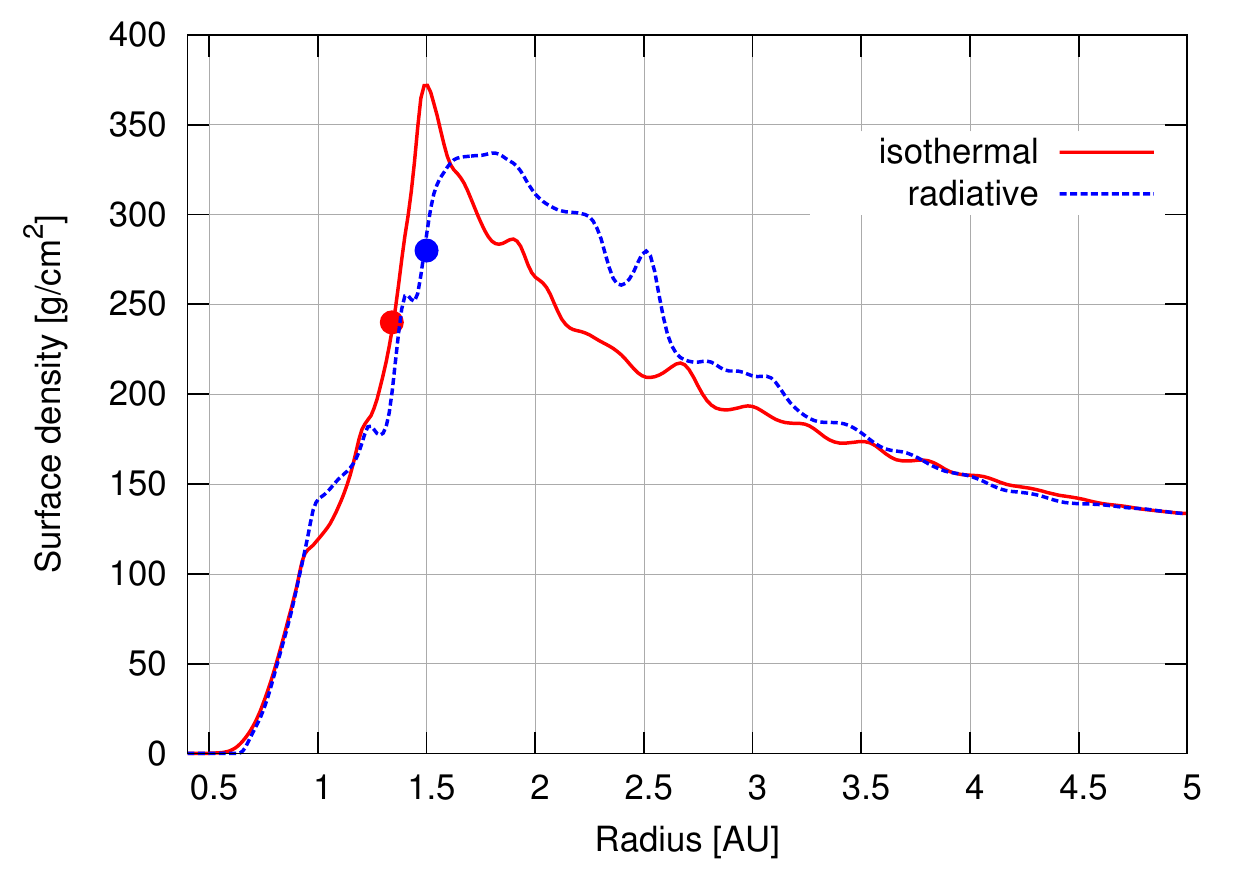} \\
\caption{Graphs of the azimuthally averaged radial surface density of the isothermal
and radiative disk models of Fig. 8 with
an embedded planet. The density distributions are taken near the final state
of the evolutions shown in Fig.~\ref{fig:k34f1p-ap-ep}.
The big colored dot in each graph represents the semi-major axis of the planet at these times. 
For illustrative purposes, we have moved the circles close to their corresponding curves.
}
\label{fig:k34f1p-sig}
\end{figure}

The final position of the planet in the two models is illustrated in Fig.~\ref{fig:k34f1p-sig} where
the radial surface density distribution is plotted together with the location of the planet.
In both cases, the planet’s final orbit is near the inner edge of the disk where the density slope is positive.
As shown here, the final position of the planet lies just inside of the peak density.
This can be compared to the case of Kepler-38 where the central binary is on an almost circular orbit 
and the planet ends up in a position closer to the inner binary where the density drops
to less than 20\% of its maximum value \citep{2014A&A...564A..72K}.

\begin{figure}
\center
\includegraphics[width=0.45\textwidth]{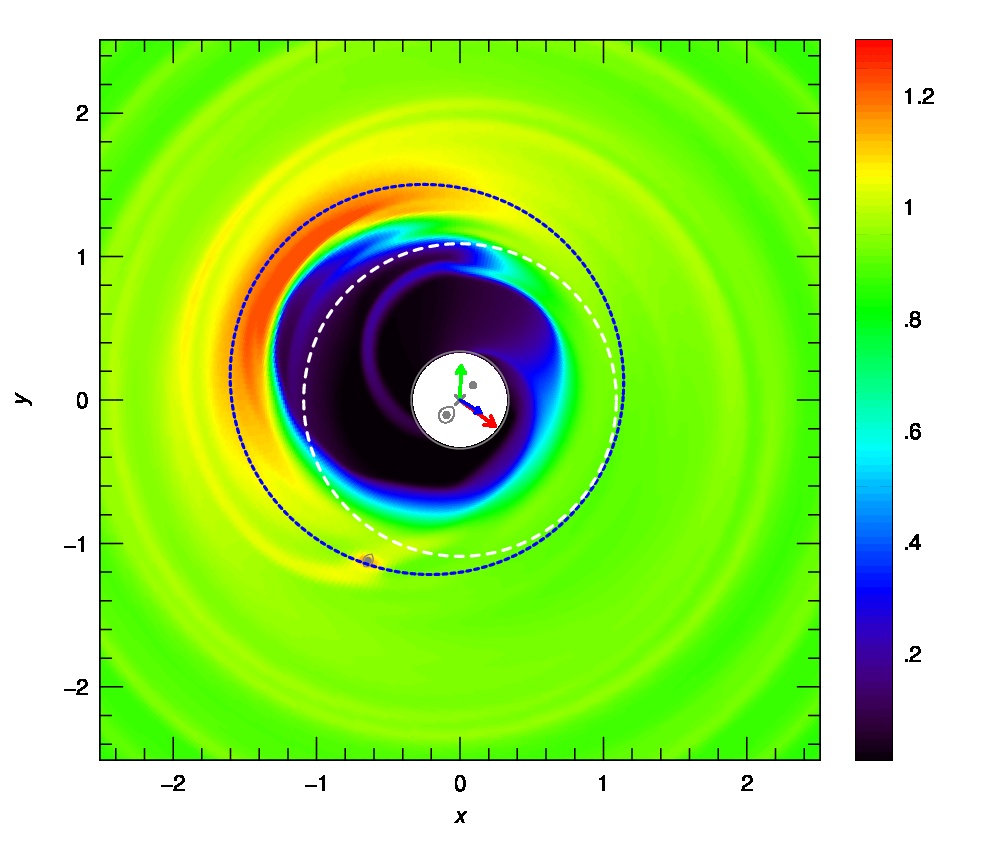} \\
\caption{2D surface density distribution for the isothermal disk model near the final state of its evolution.
The gray dots refer to the positions of the two binary stars and the planet. The Roche lobe of the system of 
the secondary star and planet are also shown. The blue dashed line refers to the orbit of the planet and the 
blue arrow points to the periapse of the planetary orbit. The red arrow points to the periapse of the inner, 
eccentric disk while the green arrow points to the periapse of the binary. The white dashed circle refers to 
the observed semi-major axis of the planet with a radius of 1.09 AU.
}
\label{fig:k34b1p-2D-obs}
\end{figure}

Fig.~\ref{fig:k34b1p-2D-obs} shows the two-dimensional density distribution of the disk near the final state of planet
migration together with the positions of the binary and planet. Also shown here are the directions of the periapses of 
the binary (green arrow), the disk (red), and the planet (blue). 
The blue dashed line refers to the orbit of the planet and the white dashed circle to the observed semi-major axis of the
planet in the Kepler-34 system. For the observed orbit, we chose to plot the semi-major axis instead of the true ellipse 
because its orientation is not constant with respect to the binary.
This figure suggests that the disk's and the planet's orbits are aligned such that their corresponding periapses always
point in the same direction. The analysis of the time evolution of the system supports this conjecture and confirms that
in the equilibrium state the planet and disk precess exactly with the same rate of 
about $3^\circ$/yr (i.e. with the same rate as the disk without the planet) see Fig.~\ref{fig:k34h-omc1}. 
We also ran a test simulation that continued this model but with a very small disk mass so that the 
planet would not feel the
disk anymore. Interestingly, in this case the planet precessed with the same rate as before, indicating that the presence 
of the disk does not have a significant influence on this process. This result seems to imply that the precession rate of the 
inner disk equals that of small massless particles, a result that will be analysed in our subsequent studies.

The strict alignment of the planetary orbit with that of the eccentric inner disk implies that
near apocenter, the planet is always positioned outside of the maximum density of the disk, as
indicated with the dotted blue line in Fig.~\ref{fig:k34b1p-2D-obs}. 
That means, a comparison between the semi-major axis of the planet and the radial location of the maximum of the azimuthally 
averaged density (see Fig.~\ref{fig:k34f1p-sig}) can give the wrong impression that the planet orbits inside
the density maximum.
 
\begin{figure}
\center
\includegraphics[width=0.45\textwidth]{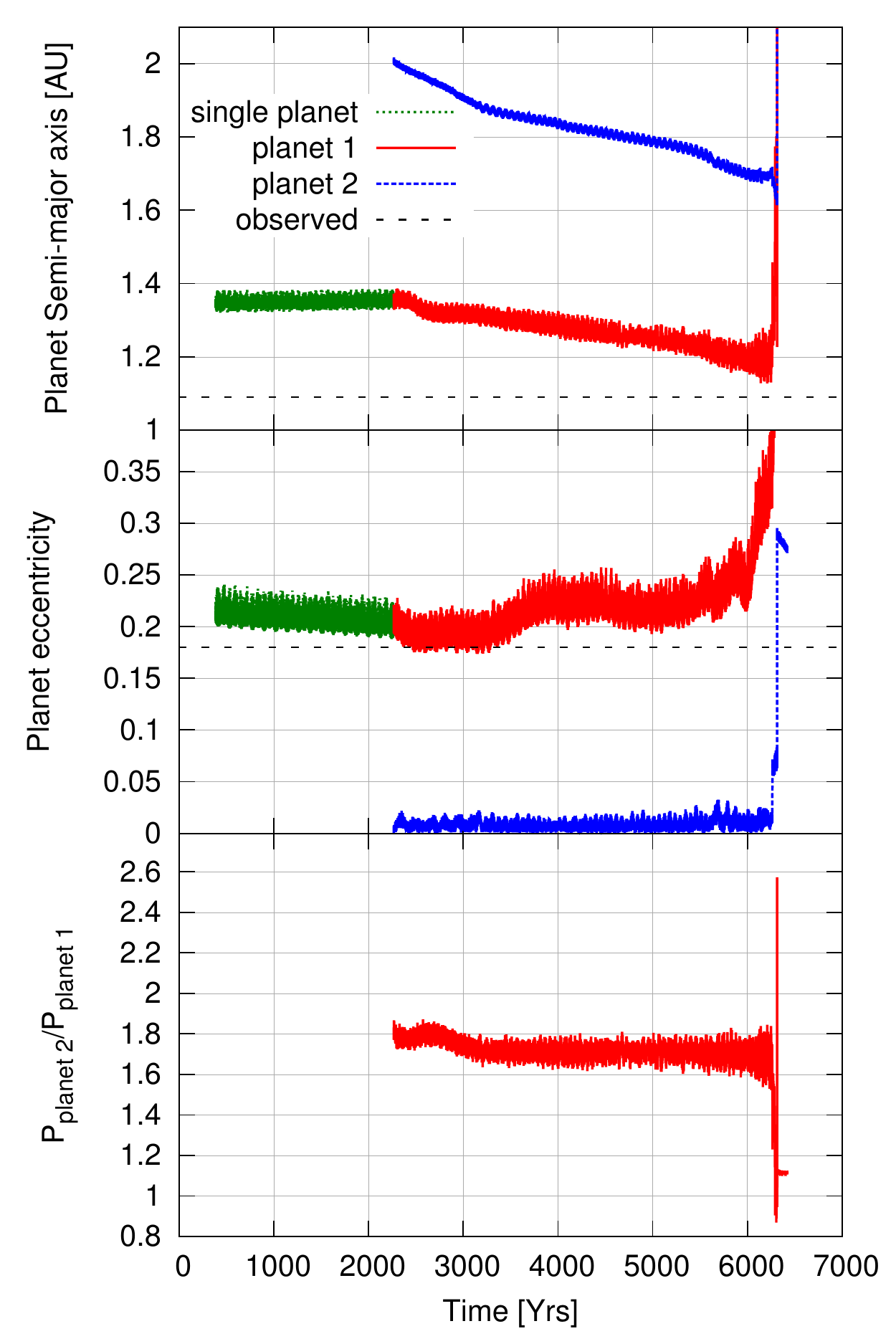} \\
\caption{The evolution of the semi-major axis (top) and eccentricity (middle) of two embedded planets.
The simulation was continued from the isothermal model shown in Fig.~\ref{fig:k34f1p-ap-ep} (red line)
by adding an additional planet with the same mass at 2.0\,AU. The green curve corresponds to the original model with the new planet
added at $t\approx 2300$\,yrs. The red curves correspond to the (original) inner planet and the blue curves is for the
(additional) outer planet. The bottom panel shows the period ratio (outer/inner) of the two planets.
}
\label{fig:k34b2p-aepratp}
\end{figure}

\section{Evolution with two planets}

As shown in previous sections, a single, embedded planet in a disk around Kepler 34 stops its migration 
well outside the observed orbit
of Kepler-34 b. This large deviation from the observed orbit is caused mainly by the wide, eccentric inner hole of the disk
that does not allow the planet to migrate closer to the star.
One possible solution to this discrepancy would be to consider an additional planet in the system located
initially at a larger distance in the disk.
During its inward migration, this outer planet, while still embedded in the disk, will interact with the inner planet 
scattering it further inward closer to the central binary.

To examine this scenario, we carried out simulations, starting from the single planet models displayed in Fig.~\ref{fig:k34f1p-ap-ep},
with an additional planet with the same mass as that of Kepler-34 b. 
This planet was added to simulations at different evolutionary times and was allowed to migrate due to its interaction with the disk.
All simulations lead to similar results. We show a sample of these simulations for an isothermal model in Fig.~\ref{fig:k34b2p-aepratp}.
In the first phase (from $t \approx 400$ to $t \approx 2300$), the final part of the
simulation of Fig.~\ref{fig:k34f1p-ap-ep} that was in red, is shown (now in green color) where we have shifted the time axis. 
Then, at $t \approx 2300$, the second planet is added. 
This new planet migrates inward (blue curve) with an initial rate similar to that of the original planet (shown in red).
At $t \approx 3200$, the migration rate suddenly changes to a slower rate. At the same time, the inner planet begins to move inward
at a similar pace, and its eccentricity is slightly increased (red curves). 
The period ratio $P_\mathrm{outer}/P_\mathrm{inner}$ (lower panel) indicates a capture
into the 7:4 resonance, which is confirmed by an analyses of the resonant angles.
During its subsequent evolution, the system remains in resonance and eventually the eccentricity of the inner planet reachers high
values and the system becomes unstable. Similar process in the radiative case of Fig.~\ref{fig:k34f1p-ap-ep} shows that in that simulation,
the two planets end up in a 3:2 resonance, and again the system becomes unstable. We also carried out
simulations where we varied the mass of the additional planet, the disk mass, and the time of second planet insertion. We found
that in general, the system goes through similar evolutionary paths: the two planets are captured in different resonances 
(2:1, 3:2 or 5:3) and they ultimately scatter each other in unstable orbits.

\begin{figure}
\center
\includegraphics[width=0.45\textwidth]{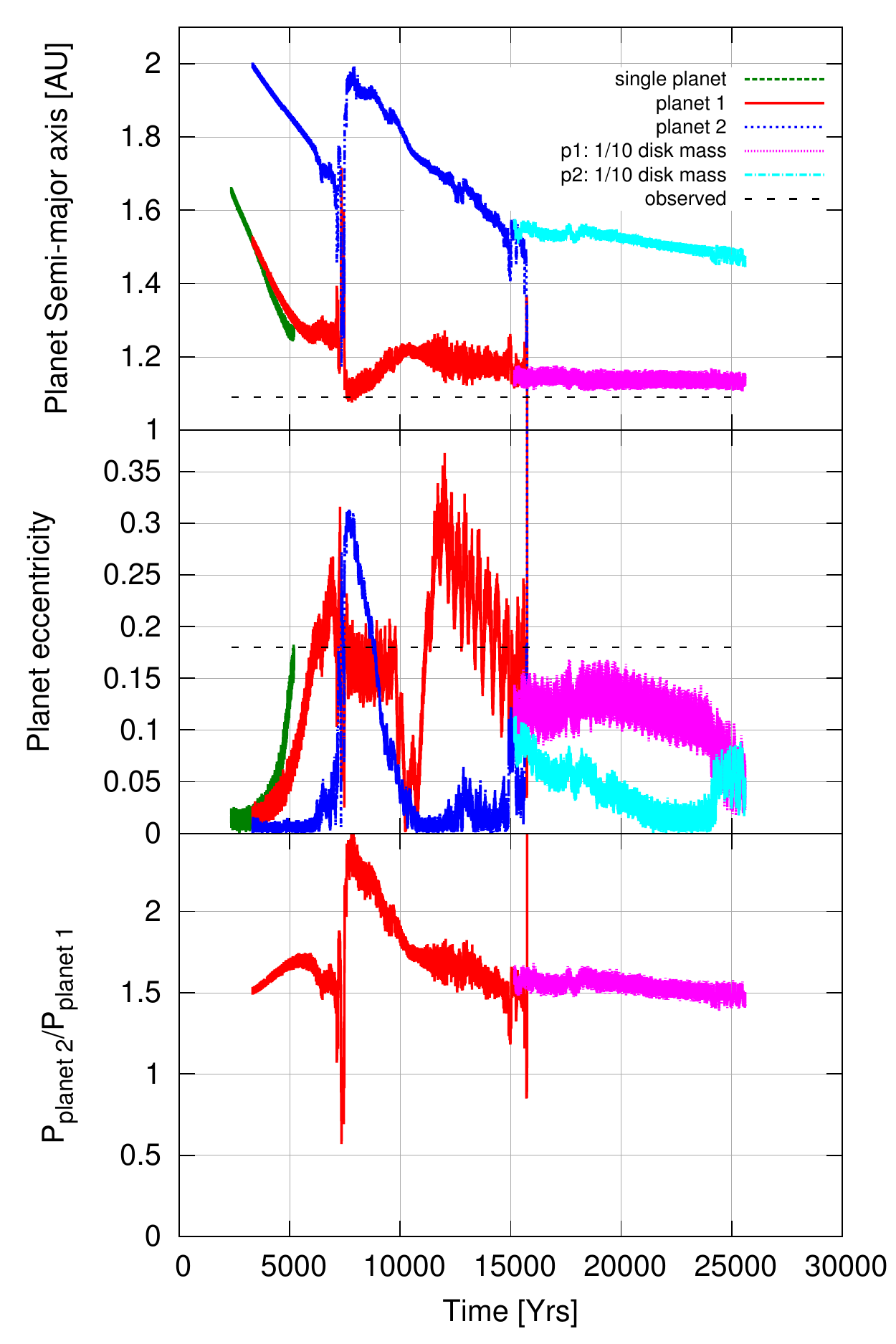} \\
\caption{The evolution of the semi-major axis (top), eccentricity (middle) and period ratios of two embedded
planets in the disk.
The simulation was continued from the isothermal model shown in Fig.~\ref{fig:k34f1p-ap-ep} (red line)
that is shown in green here. At $t \approx 3500$yr, a second planet was added (blue line) at $2.0$\,AU. At $t \approx 7000$yr, 
the planets entered a 3:2 mean-motion resonance and experienced a scattering event. 
At a later time ($t\approx 16000$), the original evolution with the two planets became eventually unstable.
At around $t\approx 15000$, the disk mass was reduced by a factor of 10, and the simulation was continued (shown in purple for the
outer planet and in light blue for the inner).
}
\label{fig:k34b6px-aepratpv}
\end{figure}

\subsection{Evolution with reduced disk mass}

In the previous section, we noted that the evolution of two planets embedded in the disk always resulted in their mutual
scattering and unstable configurations. Despite this instability, we found that after a scattering event,
one of the planets remained at a location close to the observed orbit of Kepler-34 b. This motivated us to continue the simulation
in one of this cases with a reduced disk mass to examine whether
the resulted configuration could remain stable. A lower disk mass is to be expected in the final phase of planet evolution and 
our mass reduction somewhat mimics that state of disk dissipation. 

Fig.~\ref{fig:k34b6px-aepratpv} shows the results of one of these simulations. Here, we again started the simulation from the isothermal case of
Fig.~\ref{fig:k34f1p-ap-ep}, now at an earlier time. The evolution of the initial single planet is shown again in green.
At $t \approx 3500$yr into the simulation, the second planet was added at $2.0$\,AU (blue line) and 
evolved with the original planet (now shown in red). At around $t \approx 7000$yr, the system entered a 3:2 MMR
followed, shortly, by a non-destructive scattering event where the outer planet ended up near its starting position at $2$\,AU
and the inner planet moved slightly inward, close to the observed location of Kepler-34 b.
Had we stopped the simulation at this point, we would have had a perfect match with the observations for the inner
planet. However, the subsequent evolution (with the disk still being present) lead to an inward migration of the outer planet, and
a slight outward motion of the inner planet (see appendix \ref{app:outward}). At around $t \approx 15000$, the planets were
captured again into the 3:2 resonance which at $t\approx 16000$, resulted in a violent scattering event.

To study the effect of disk dissipation, we restarted the above-mentioned simulation just before the final scattering event and
after $t \approx 15000$, with 1/10 of the original disk mass. For this phase, the evolution of the outer planet is shown 
in light blue and that of the inner planet is in purple. The bottom panel shows the period ratio in purple as well.
During this low disk mass phase, the inward migration of the outer planet continues with a very small rate, 
while the inner planet remains at its location. Eventually the planets are captured in a 3:2 MMR, however, the system remains
stable as is indicated by the small values of the planets’ eccentricities. 

\section{Summary and discussion}

Using 2D hydrodynamical simulations, we studied the evolution of planets embedded in a circumbinary disk
for the parameters of the Kepler-34 system. The stellar binary consists of two stars of
nearly equal masses ($\sim 1 {M_\odot}$) orbiting each other in a relatively high
eccentric orbit with $e_\mathrm{bin} = 0.52$.
These orbital characteristics of Kepler-34 make this system much more dynamic than some of the other systems that host circumbinary planets.

Our investigation followed that of Kepler-38 \citep{2014A&A...564A..72K} where we adopted a two-step approach.
We first studied the structure of a circumbinary disk without 
a planet, and then included a planet in the disk and followed its subsequent evolution.
We considered locally isothermal disks as well as more realistic disks that include
viscous dissipation, vertical cooling, and radiative diffusion. In the following, we briefly present 
our most important results.

As shown in Fig.~\ref{fig:k34a-sigecc}, a highly eccentric stellar binary generates a large inner hole in the disk with a high eccentricity.
As a result, around an eccentric binary such as Kepler-34, the maximum value of the density distribution lies at farther distances
(around 7 $a_\mathrm{bin}$) compared to that in systems with circular binaries such Kepler-38 (at about 5  $a_\mathrm{bin}$).
Our simulations also showed that the inner region of the disk will precess in a prograde sense. For the system of Kepler-34,
the rate of this precession is about $\varpi_\mathrm{disk} = 3^\circ$/yr. 
In turn, the disk that has a total mass of 0.015 $M_\odot$ within 5\,AU (our reference model),
induces a slow prograde precession on the binary with a rate of $\varpi_\mathrm{bin} = 0.05^\circ$/yr.

To study the evolution of an embedded planet in a disk around Kepler-34, we added a planet with a mass of 
$0.22 M_\mathrm{Jup}$ (the observed value of the mass of Kepler-34 b) to the system and
allowed the planet to migrate in the disk as a result of disk torques acting on it.
In our reference disk model, the planet migration was so rapid that the its orbit became unstable by the time 
it had reached a distance of about 1.35\,AU from the barycenter of the binary. When reducing the mass of the disk to 25\%,
and considering a lower viscosity ($\alpha = 0.004$), the orbit of the planet became stable.
Here, isothermal and radiative models resulted in qualitatively similar evolutions. The planet migrated
inward to a distance of about 1.3 \,AU from the center and then moved out to settle into an equilibrium state
of about 1.35\,AU for the isothermal and 1.5\,AU for the radiative models.
A test simulation, in which the planet started inside the hole of the disk, at the observed distance of $1.09$AU,
showed that the planet moves outward and settles approximately at the same location as well
(see Appendix).

Results of our simulations indicated that the final location of the planet is robust and determined by the size and 
shape (eccentricity) of the disk inner hole.
The final distance of the planet from the binary's barycenter is larger than the observed semi-major axis of Kepler-34 b, however 
its eccentricity is in good agreement with its observed value. In this final state, the precession rate of the planet orbit 
becomes equal to the disk precession rate where they enter in a state of apsidal corotation with an alignment
of their periapses. As a result, the planet always orbits outside the disk’s eccentric hole, avoiding moving further into the disk.
These results seem to indicate that in an eccentric binary such as Kepler-34, the evolution of the disk makes it difficult
for the planet to reach close distances from the binary. The final position of the planet is determined by the size of the disk inner hole
which becomes larger for eccentric binaries. 

The difficulty in reaching close orbits for a planet in an eccentric disk combined with 
the discovery of the circumbinary planetary system of Kepler-47 with multiple planets motivated us to assume
that Kepler-34 may contain (or might have contained) at least one additional planet.
To investigate the evolution of multiple planets in the disk, we added an additional planet, in a exterior orbit, 
to our single-planet disk models and followed the combined evolution of both planets.
For isothermal and radiative disks, these typically resulted in a capture of the two planets in 
low-order, mean-motion resonances with period ratios of 2:1, 3:2, 5:3 and 7:4. In all cases,
the inward migration of the resonantly coupled pair resulted in unstable configurations when
the inner planet reached a distance of about 1.1 to 1.2\,AU from the binary's barycenter. In these simulations,
one planet was typically ejected from
the system and the second one was scattered into an orbit with much larger semi-major axis.

Upon reducing the disk mass by a factor of 10, we found that the planet orbits could stabilize with their subsequent
evolution resulting in a system where the inner planet has a semimajor axis close to its observed value. 
This suggested that a model in which a pair of planets are driven toward the binary just before disk dispersal
represent a possible scenario for explaining the orbital configuration of the Kepler-34 system. In this case,
it may be interesting to search for additional planets in the system.

As we have shown in this study, the evolution of planets in a disk depends crucially on the structure of the inner hole of the disk
which itself depends on the binary eccentricity.
The detailed dependence on disk parameter such as viscosity and disk mass will be investigated in more detail
in future studies.

We note that in the models presented in this paper, we restricted ourselves to flat two-dimensional disks. Given the flatness of the 
observed circumbinary systems this is not an unrealistic assumption. However, a more comprehensive study of the dynamical evolution of these
systems requires full three-dimensional simulations of radiative disks around binary star systems.

\begin{acknowledgements}
N.H. acknowledges support from the NASA ADAP grant NNX13AF20G, and the Alexander von Humboldt Foundation.
N.H. would also like to thank the Computational Physics Group at the University of T\"ubingen for their kind
hospitality during the course of this project.

\end{acknowledgements}
\bibliography{kep34-refs}{}
\bibliographystyle{aa}

\newpage 

\appendix 

\section{Evolution of a close in planet}
\label{app:outward}
\begin{figure}[h]
\center
\includegraphics[width=0.45\textwidth]{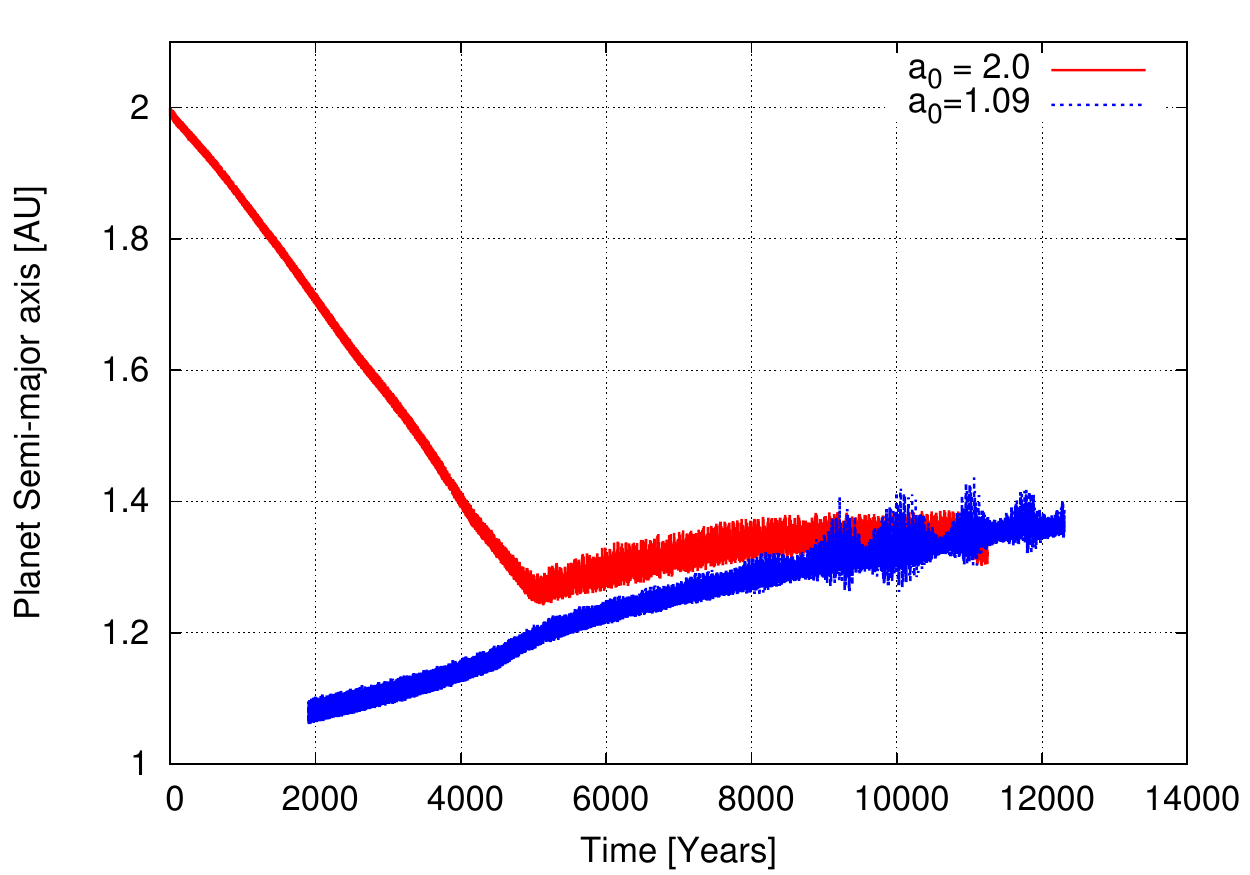} \\
\caption{The evolution of the semi-major axis of two the planets in an isothermal disk.
The red curve corresponds to the model displayed in Fig.~\ref{fig:k34f1p-ap-ep}.
The blue curve corresponds to a planet that has started with an initial position
of $a_0 = 1.09$ inside the inner hole of the disk.
}
\label{fig:k34b8p-ap2}
\end{figure}

To test the robustness of the final position of a migrating planet in an eccentric disk around
the Kepler-34 binary, we performed additional simulations where the planet was started {\it inside} the
inner hole of the circumbinary disk.
In Fig.~\ref{fig:k34b8p-ap2}, we show the evolution of a planet in two systems. In the first system shown in red,
the planet starts at $r_0= 2.0$\,AU as displayed already in Fig.~\ref{fig:k34f1p-ap-ep}.
In the second case, shown in blue, the planet starts at $r_0 = 1.09$\,AU inside
the hole of the disk. In both cases, the initial eccentricity of the planet was set to $e_0 =0$. In the second
system, the starting position of the planet lies at the present observed location of Kepler-34 b.

The simulations show that the planet starting inside the inner hole of the disk does not remain at its orbit
but moves slowly outward.
In both systems, the final position of the planet is the same, at $\approx$ 1.35\,AU from the binary barycenter.
The inner planet that moves out shows large oscillations in its eccentricity causing variations to appear in its semi-major axis.
When running simulations for long times, these variations seem to become smaller. The outward migration in the second system 
can be understood by the interaction of an eccentric disk with the orbiting planet. 
During one orbit of the planet, because of the oscillations in its semi-major axis, it enters periodically into the eccentric 
disk where is experiences a positive net torque that drives the planets outward.

\end{document}